\documentclass[english]{spie}
\usepackage[T1]{fontenc}
\usepackage[latin9]{inputenc}
\usepackage{amsmath}
\usepackage{amssymb}
\usepackage{graphicx}

\makeatletter
\usepackage{epstopdf}\usepackage{epsfig}

\usepackage{babel}

\makeatother

\usepackage{babel}
\begin{document}

\title{Quantum Zeno and anti-Zeno effects in an asymmetric nonlinear optical
coupler}

\author{Kishore Thapliyal{*}, Anirban Pathak\\
 Jaypee Institute of Information Technology, A-10, Sector-62, Noida,
UP-201307, India}

\authorinfo{{*}tkishore36@yahoo.com}
\maketitle
\begin{abstract}
Quantum Zeno and anti-Zeno effects in an asymmetric nonlinear optical
coupler are studied. The asymmetric nonlinear optical coupler is composed
of a linear waveguide ($\chi^{\left(1\right)}$) and a nonlinear waveguide
($\chi^{\left(2\right)}$) interacting with each other through the
evanescent waves. The nonlinear waveguide has quadratic nonlinearity
and it operates under second harmonic generation. A completely quantum
mechanical description is used to describe the system. The closed
form analytic solutions of Heisenberg's equations of motion for the
different field modes are obtained using Sen-Mandal perturbative approach.
In the coupler, the linear waveguide acts as a probe on the system
(nonlinear waveguide). The effect of the presence of the probe (linear
waveguide) on the photon statistics of the second harmonic mode of
the system is considered as quantum Zeno and anti-Zeno effects. Further,
it is also shown that in the stimulated case, it is easy to switch
between quantum Zeno and anti-Zeno effects just by controlling the
phase of the second harmonic mode of the asymmetric coupler. 
\end{abstract}


\section{INTRODUCTION}

Zeno's paradoxes have been in discussion since fifth century. In 1977,
Mishra and Sudarshan \cite{misra1977zeno} introduced a quantum analogue
of Zeno's paradox, which is later termed as quantum Zeno effect. Quantum
Zeno effect in the original formulation refers to the inhibition of
the temporal evolution of a system on continuous measurement \cite{misra1977zeno,Peres-zeno,Khalfin},
while quantum anti-Zeno or inverse Zeno effect refers to the enhancement
of the evolution instead of the inhibition \cite{anti-zeno} (see
\cite{venugopalan2007zeno_review,Saverio-rev} for reviews on Zeno
and anti-Zeno effect). In the last four decades, quantum Zeno effect
has been studied in different physical systems, such as two coupled
nonlinear optical processes\cite{2nonlinear}, parametric down-conversion\cite{perina-zeno}
and cascaded parametric down-conversion with losses\cite{Zeno-with-loss}.
Similarly, quantum anti-Zeno effect is also reported in various physical
systems. Specifically, quantum anti-Zeno effect is observed in parametric
down-conversion\cite{anti-zeno-downconv} and in radioactive decay
processes, where the measurement causes the system to disintegrate
\cite{exp-anti-zeno}. In Ref. \cite{zeno-anti-zeno-2level}, both
the effects (quantum Zeno and anti-Zeno) are reported in two-level
systems. Further, a geometrical criterion for transition between the
Zeno and anti-Zeno effects has also been discussed in Ref. \cite{Inverse-zeno}.
Quantum Zeno effect using environment-induced decoherence theory has
also been discussed in the past\cite{Zeno+decohrence}. Agarwal and
Tewari proposed a scheme for an all optical realization of quantum
Zeno effect using an arrangement of beam splitters \cite{agarwal}.
In addition to the above mentioned theoretical studies, quantum Zeno
effect has also been experimentally realized in trapped beryllium
ions \cite{exp-zeno} and with the help of rotators and polarizers
\cite{Exp2-Zeno}. \\

Recently, the interest on quantum Zeno effect has increased by manifold
as it has found its applications in counterfactual quantum computation\cite{qu-interr},
where computation is accomplished using the computer in superposition
of running and not running states and later to infer the solution
from it; counterfactual quantum communication, in which information
is sent without sending the information encoded particles through
the communication channel\cite{salih-counterfactual}; and quantum
Zeno tomography, which essentially uses interaction free measurement\cite{tomography1}.
Interestingly, detection of an absorbing object without any interaction
with light, using quantum Zeno effect with higher efficiency has already
been demonstrated \cite{interaction-free-measurement}. These facts
motivated the study of Zeno effect in macroscopic systems\cite{macroscopic-zeno},
too. Quantum Zeno and anti-Zeno effects in the nonlinear optical couplers
have been studied in the recent past\cite{thun2002zeno-raman,rehacek2000zeno-coupler,facchi2001zeno-review,chi2-chi2-zeno}
by considering that one of the mode in the nonlinear waveguide is
coupled with the auxiliary mode in a linear waveguide, and the auxiliary
linear mode acts as a probe (continuous observation) on the evolution
of the system (nonlinear waveguide) and changes the photon statistics
of the other modes of the nonlinear waveguide, which are not coupled
with the probe. This kind of continuous interaction with an external
system is equivalent to the original inhibition or enhancement of
evolution of the system on continuous measurement as measurement is
strong coupling with a measuring device\cite{cont-int,continuous-int,continuous-meas}.\\

An important example of nonlinear optical coupler is an asymmetric
nonlinear optical coupler, consisting of a nonlinear waveguide with
$\chi^{(2)}$ nonlinearity operating under second harmonic generation
coupled with a linear waveguide. This system is studied earlier and
the nonclassical properties (such as squeezing, antibunching and entanglement)
have been reported in this system for both codirectional and contradirectional
propagation of the field in the linear waveguide (\cite{mandal2004co-coupler,kishore2014co-coupler,kishore2014contra,perina1995photon}
and references therein). We will consider here the codirectional propagation
of the fields as in contradirectional propagation the solution was
only valid at both the ends of the coupler, i.e., not valid for $0<z<L$,
where $L$ is the interaction length of the coupler \cite{kishore2014contra,perina1995photon}.
It has already been established that the nonclassical effects can
be observed in the asymmetric nonlinear coupler\cite{mandal2004co-coupler,kishore2014co-coupler,kishore2014contra,perina1995photon}.
The presence of quantum Zeno and anti-Zeno effects reported in this
work further establishes the existence of nonclassicality in this
asymmetric coupler.\\

To study the quantum Zeno and anti-Zeno effects closed form analytic
expressions for different field operators are used here. These expressions
were obtained earlier by using Sen-Mandal perturbative approach\cite{mandal2004co-coupler,kishore2014co-coupler}
and a completely quantum mechanical description of the system. The
solutions of Heisenberg's equations of motion used here are better
than the conventional short-length solutions as these solutions are
not restricted by length\cite{perina1995photon}. In what follows,
we use the solutions reported in Refs. \cite{mandal2004co-coupler,kishore2014co-coupler}
to establish the existence of quantum Zeno and anti-Zeno effects in
the asymmetric nonlinear optical coupler.\\

The remaining part of this paper is organized as follows. In Section
2, we briefly describe the momentum operator that provides a completely
quantum mechanical description of the system and also note the analytic
expressions of the field operators required for the present study.
In Section 3, we show the existence of quantum Zeno and anti-Zeno
effects in the asymmetric nonlinear optical coupler and show the spatial
evolution of the Zeno parameter. The variation of the Zeno parameter
has also been studied with the phase of the coherent input in the
second harmonic mode of the nonlinear waveguide, phase mismatch between
fundamental and second harmonic modes in the nonlinear waveguide,
and linear coupling between probe and the system. Finally, the paper
is concluded in Section 4.

\section{THE SYSTEM AND SOLUTION\label{sec:sys-sol}}

\begin{figure}
\begin{centering}
\includegraphics[scale=0.8]{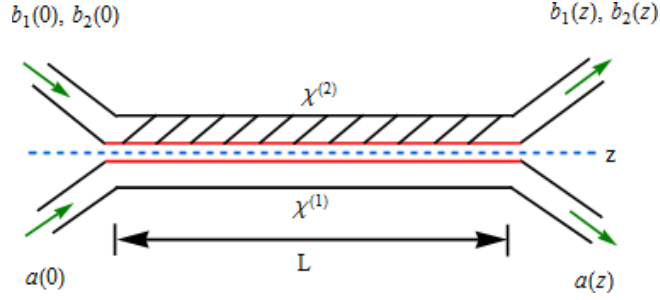} 
\par\end{centering}

\protect\caption{\label{fig:Schematic-diagram}(Color online) Schematic diagram of
an asymmetric nonlinear optical coupler of interaction length $L$
in codirectional propagation prepared by combining a linear waveguide
with a nonlinear (quadratic) waveguide operated by second harmonic
generation.}
\end{figure}

The asymmetric nonlinear optical coupler, schematically shown in Fig.
\ref{fig:Schematic-diagram}, is prepared by combining a linear waveguide
with a nonlinear waveguide of $\chi^{(2)}$ nonlinearity. As $\chi^{(2)}$
medium can produce second harmonic generation, we may say that the
codirectional asymmetric nonlinear optical coupler studied here for
the investigation of possibility of observing quantum Zeno and anti-Zeno
effects is operated under second harmonic generation. The momentum
operator for this specific coupler in interaction picture is\cite{mandal2004co-coupler}
\begin{equation}
G=-\hbar kab_{1}^{\dagger}-\hbar\Gamma b_{1}^{2}b_{2}^{\dagger}\exp(i\Delta kz)\,+{\rm H.c}.\,,\label{eq:hamiltonian}
\end{equation}
where ${\rm H.c.}$ stands for the Hermitian conjugate, and $\Delta k=|2k_{1}-k_{2}|$
is the phase mismatch between the fundamental and second harmonic
beams, and $k$ ($\Gamma$) being the linear (nonlinear) coupling
constant is proportional to the linear (nonlinear) susceptibility
$\chi^{(1)}$ $(\chi^{(2)})$. The value of $\chi^{(2)}$ is considerably
smaller than $\chi^{(1)}$ (typically $\chi^{(2)}/\chi^{(1)}\,\simeq10^{-6})$
which leads to $\Gamma\ll k$, unless an extremely strong pump is
present in the nonlinear waveguide.\\

In Refs. \cite{mandal2004co-coupler,kishore2014co-coupler} closed
form analytic expressions for the evolution of the field operators
corresponding to the Hamiltonian (\ref{eq:hamiltonian}) were obtained
using Sen-Mandal perturbative approach valid up to the linear power
of the nonlinear coupling coefficient $\Gamma$. The field opeartors
reported there are: 
\begin{equation}
\begin{array}{lcl}
a(z) & = & f_{1}a(0)+f_{2}b_{1}(0)+f_{3}b_{1}^{\dagger}(0)b_{2}(0)+f_{4}a^{\dagger}(0)b_{2}(0),\\
b_{1}(z) & = & g_{1}a(0)+g_{2}b_{1}(0)+g_{3}b_{1}^{\dagger}(0)b_{2}(0)+g_{4}a^{\dagger}(0)b_{2}(0),\\
b_{2}(z) & = & h_{1}b_{2}(0)+h_{2}b_{1}^{2}(0)+h_{3}b_{1}(0)a(0)+h_{4}a^{2}(0),
\end{array}\label{eq:assumed-sol}
\end{equation}
where 
\begin{equation}
\begin{array}{lcl}
f_{1} & = & g_{2}=\cos|k|z,\\
f_{2} & = & -g_{1}*=-\frac{ik^{*}}{|k|}\sin|k|z,\\
f_{3} & = & \frac{2k^{*}\Gamma^{*}}{4|k|^{2}-(\Delta k)^{2}}\left[G_{-}f_{1}+\frac{f_{2}}{k^{*}}\left\{ \Delta k-\frac{2|k|^{2}}{\Delta k}G_{-}\right\} \right],\\
f_{4} & = & \frac{4k^{*2}\Gamma^{*}}{\Delta k\left[4|k|^{2}-(\Delta k)^{2}\right]}G_{-}f_{1}+\frac{2k^{*}\Gamma^{*}}{\left[4|k|^{2}-(\Delta k)^{2}\right]}G_{+}f_{2},\\
g_{3} & = & \frac{2\Gamma^{*}k}{\left[4|k|^{2}-(\Delta k)^{2}\right]}G_{+}f_{2}-\frac{2\Gamma^{*}\left(2|k|^{2}-(\Delta k)^{2}\right)f_{1}}{\Delta k\left[4|k|^{2}-(\Delta k)^{2}\right]}G_{-},\\
g_{4} & = & \frac{4\Gamma^{*}|k|^{2}}{\Delta k\left[4|k|^{2}-(\Delta k)^{2}\right]}f_{2}-\frac{2\Gamma^{*}\left(2|k|^{2}-(\Delta k)^{2}\right)}{\Delta k\left[4|k|^{2}-(\Delta k)^{2}\right]}\left(G_{+}-1\right)f_{2}+\frac{2k^{*}\Gamma^{*}}{\left[4|k|^{2}-(\Delta k)^{2}\right]}G_{-}f_{1},\\
h_{1} & = & 1,\\
h_{2} & = & \frac{\Gamma G_{-}^{*}}{2\Delta k}-\frac{i\Gamma}{2\left[4|k|^{2}-(\Delta k)^{2}\right]}\left\{ 2|k|\left(G_{+}^{*}-1\right)\sin2|k|z-i\Delta k\left[1-\left(G_{+}^{*}-1\right)\cos2|k|z\right]\right\} ,\\
h_{3} & = & \frac{-\Gamma|k|}{k^{*}\left[4|k|^{2}-(\Delta k)^{2}\right]}\left\{ i\Delta k\left(G_{+}^{*}-1\right)\sin2|k|z+2|k|\left[1-\left(G_{+}^{*}-1\right)\cos2|k|z\right]\right\} ,\\
h_{4} & =- & \frac{\Gamma|k|^{2}G_{-}^{*}}{2k^{*^{2}}\Delta k}-\frac{i\Gamma|k|^{2}}{2k^{*^{2}}\left[4|k|^{2}-(\Delta k)^{2}\right]}\left\{ 2|k|\left(G_{+}^{*}-1\right)\sin2|k|z-i\Delta k\left[1-\left(G_{+}^{*}-1\right)\cos2|k|z\right]\right\} ,
\end{array}\label{eq:sol-coeff}
\end{equation}
where $G_{\pm}=\left[1\pm\exp(-i\Delta kz)\right].$

The number operator for the second harmonic field mode in the nonlinear
waveguide, i.e., $b_{2}$ mode is given by

\begin{equation}
N_{b_{2}}\left(z\right)=b_{2}^{\dagger}\left(z\right)b_{2}\left(z\right)=b_{2}^{\dagger}(0)b_{2}(0)+\left[h_{2}b_{2}^{\dagger}(0)b_{1}^{2}(0)+h_{3}b_{2}^{\dagger}(0)b_{1}(0)a(0)+h_{4}b_{2}^{\dagger}(0)a^{2}(0)+{\rm H.c.}\right].\label{eq:nb2}
\end{equation}
The initial state being the multimode coherent state $|\alpha\rangle|\beta\rangle|\gamma\rangle$
with all three modes $|\alpha\rangle,\,|\beta\rangle$ and $|\gamma\rangle$
the eigen kets of annihilation operators $a$, $b_{1}$ and $b_{2}$,
respectively. For example, after the operation of the field operator
$b_{2}(0)$ on such a multimode coherent state we would obtain 
\begin{equation}
b_{2}(0)|\alpha\rangle|\beta\rangle|\gamma\rangle=\gamma|\alpha\rangle|\beta\rangle|\gamma\rangle,\label{eq:coherent-state}
\end{equation}
where $|\alpha|^{2},\,|\beta|^{2}\,{\rm and}\,|\gamma|^{2}$ are the
initial number of photons in the field modes $a$, $b_{1}$ and $b_{2}$,
respectively. Further, the coupler discussed here can operate under
two conditions: spontaneous and stimulated. In the spontaneous case,
$\left|\alpha\right|\neq0,$ $\left|\beta\right|\neq0,$ and $\left|\gamma\right|=0$,
whereas in the stimulated case, $\left|\alpha\right|\neq0,$ $\left|\beta\right|\neq0,$
and $\left|\gamma\right|\neq0$.

\section{QUANTUM ZENO AND ANTI-ZENO EFFECTS\label{sec:Q-Z-ai-Zeno}}

\begin{figure}
\begin{centering}
\includegraphics[scale=0.6]{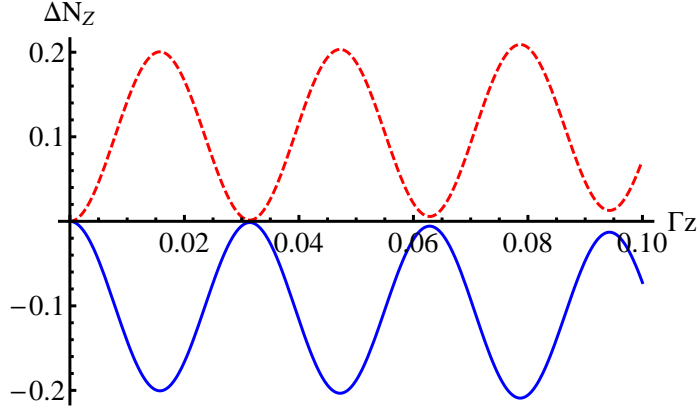} 
\par\end{centering}

\protect\caption{\label{fig:zeno}(Color online) The spatial variation of the Zeno
parameter ($\Delta N_{Z}$) in the second harmonic ($b_{2}$) mode
is shown with rescaled length ($\Gamma z$) for the initial state
$|\alpha\rangle|\beta\rangle|\gamma\rangle$ and $k=0.1,\,\Gamma=0.001,\,\Delta k=10^{-4},\,\alpha=5$
and $\beta=2,$ where the smooth (blue) and dashed (red) lines correspond
to $\gamma=1$ (quantum Zeno effect) and $\gamma=-1$ (quantum anti-Zeno
effect), respectively.}
\end{figure}

The number of photons in the second harmonic mode for the initial
multimode coherent state (\ref{eq:coherent-state}) is given by

\begin{equation}
\left\langle N_{b_{2}}\left(z\right)\right\rangle =\left|\gamma\right|^{2}+\left[h_{2}\beta^{2}\gamma^{*}+h_{3}\alpha\beta\gamma^{*}+h_{4}\alpha^{2}\gamma^{*}+{\rm c.c.}\right].\label{eq:nb2-k}
\end{equation}
In the absence of the probe mode, i.e., $k=0$ and $\alpha=0$, we
have 
\begin{equation}
\left\langle N_{b_{2}}\left(z\right)\right\rangle _{k=0}=\left|\gamma\right|^{2}+\left[h_{2}^{\prime}\beta^{2}\gamma^{*}+{\rm c.c.}\right],\label{eq:nb2-k-0}
\end{equation}
where $h_{2}^{\prime}=h_{2}\left(k=0\right)=\frac{\Gamma}{\Delta k}\left[1-\exp(i\Delta kz)\right]$.
The effect of the presence of the probe mode can be given as $\left(\Delta N_{Z}=\left\langle N_{b_{2}}\left(z\right)\right\rangle -\left\langle N_{b_{2}}\left(z\right)\right\rangle _{k=0}\right)$,
where $\Delta N_{Z}$ is the Zeno parameter. The non-zero value of
the Zeno parameter implies that the presence of the probe affects
the evolution of the photon statistics of the system, i.e., the positive
(negative) value of the Zeno parameter means the photon generation
is increased (decreased) due to the continuous measurement of the
probe on the linear mode of the nonlinear waveguide. In the present
case, using Eqs. (\ref{eq:nb2-k}) and (\ref{eq:nb2-k-0}), we may
obtain the analytic expression for the Zeno parameter as 
\begin{equation}
\Delta N_{Z}=\left[\left\{ \left(h_{2}-h_{2}^{\prime}\right)\beta^{2}+h_{3}\alpha\beta+h_{4}\alpha^{2}\right\} \gamma^{*}+{\rm c.c.}\right].\label{eq:zeno}
\end{equation}
When the Zeno parameter becomes negative (positive), it implies the
existence of quantum Zeno (anti-Zeno) effect. Here, the expression
for the Zeno parameter is of the form $\left|\gamma\right|F\left(h_{i},\phi\right)$,
where we have considered $\gamma\equiv\left|\gamma\right|\exp\left(i\phi\right)$.
So, the Zeno parameter becomes zero in the spontaneous case, and neither
the quantum Zeno effect nor the anti-Zeno effect can be observed.
In the stimulated case, the functional form of the Zeno parameter
suggests that we can control the quantum Zeno and anti-Zeno effects
in the asymmetric coupler just by controlling the phase of the input
coherent beam of the $b_{2}$ mode, i.e., changing $\phi$. As the
phase change of $\pi$ in $\gamma$ changes the quantum Zeno effect
into quantum anti-Zeno effect, and vice versa. For example, in Fig.
\ref{fig:zeno}, we show the spatial evolution of the Zeno parameter
for $\phi=0$ (smooth blue line) and $\phi=\pi$ (dashed red line).
While the choice $\phi=0$ illustrates the existence of quantum Zeno
effect, $\phi=\pi$ is found to illustrate the existence of quantum
anti-Zeno effect.

\begin{figure}
\begin{centering}
\includegraphics[scale=0.8]{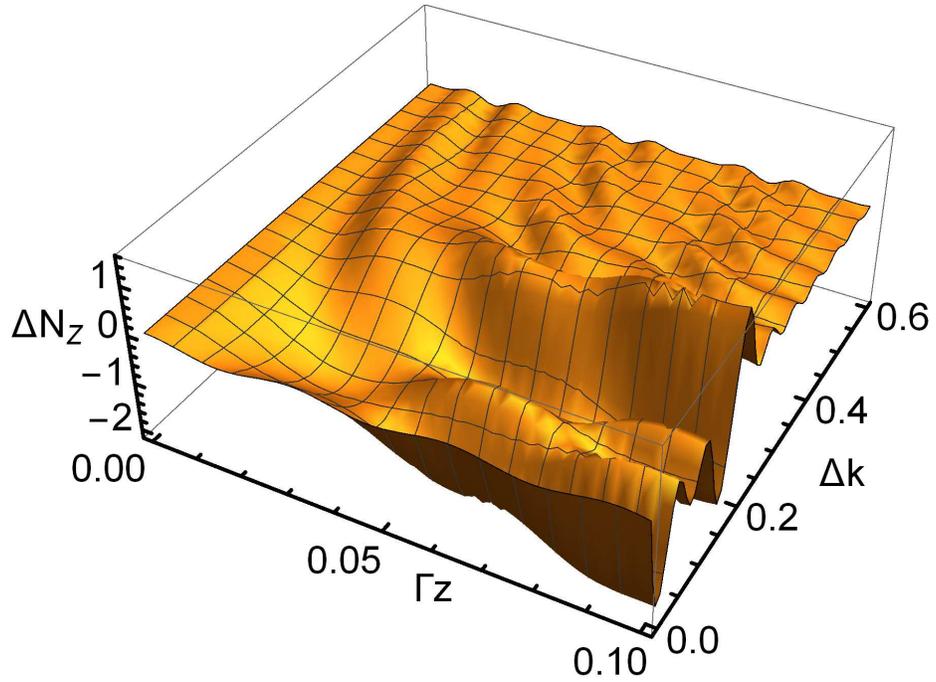} 
\par\end{centering}

\protect\caption{\label{fig:zeno-mismatch}(Color online) The spatial variation of
the Zeno parameter ($\Delta N_{Z}$) in the second harmonic ($b_{2}$)
mode with rescaled length ($\Gamma z$) and phase mismatch ($\Delta k$)
between fundamental and second harmonic mode in the nonlinear waveguide
for the initial state $|\alpha\rangle|\beta\rangle|\gamma\rangle$
and $k=0.1,\,\Gamma=0.001,\,\alpha=5$ and $\beta=2,$ and $\gamma=1$.
Transition between quantum Zeno and anti-Zeno effects with change
in phase mismatch can be observed.}
\end{figure}

\begin{figure}
\begin{centering}
\includegraphics[scale=0.8]{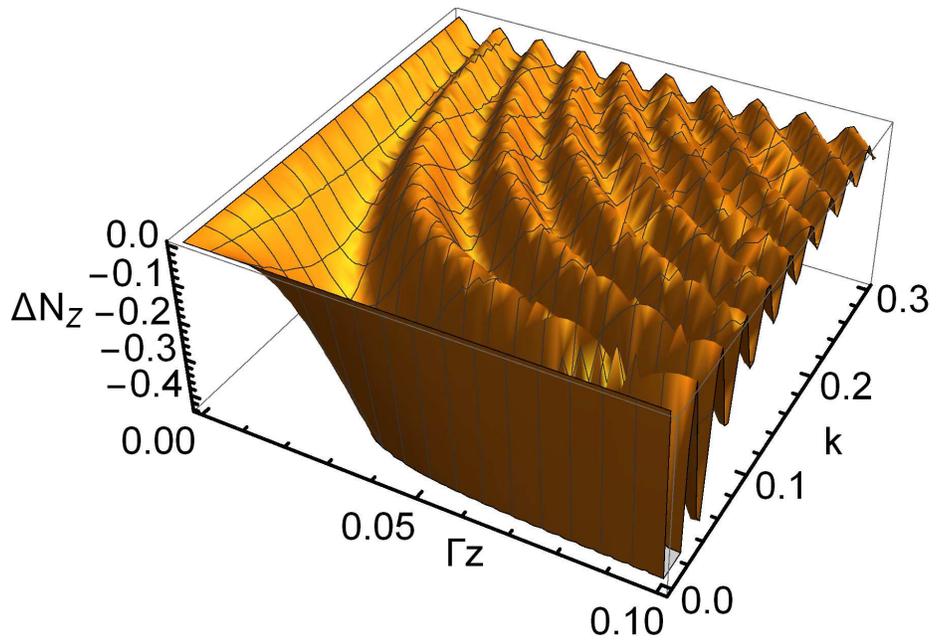} 
\par\end{centering}

\protect\caption{\label{fig:zeno-k}(Color online) The spatial variation of the Zeno
parameter ($\Delta N_{Z}$) depicts quantum Zeno effect in the second
harmonic ($b_{2}$) mode is shown with rescaled length ($\Gamma z$)
and linear coupling ($k$) between the waveguides for the initial
state $|\alpha\rangle|\beta\rangle|\gamma\rangle$ and $\Gamma=0.001,\,\Delta k=10^{-4},\,\alpha=5$
and $\beta=2,$ and $\gamma=1$.}
\end{figure}

Further, the effect of change of the other parameters on the Zeno
parameter can also be observed. Specifically, Fig. \ref{fig:zeno-mismatch}
shows that the transition between quantum Zeno and anti-Zeno effects
can be observed with change in phase mismatch between fundamental
and second harmonic modes in the nonlinear waveguide. Similarly, Fig.
\ref{fig:zeno-k} illustrates the variation of quantum Zeno effect,
as the Zeno parameter remains negative, with linear coupling between
probe and the system. A similar behaviour can also be observed for
quantum anti-Zeno effect. However, the variation of the Zeno parameter
with change in nonlinear coupling constant is observed to be linear
in nature and decreasing with nonlinear coupling constant (corresponding
plot is not included in this paper).

\section{CONCLUSION\label{sec:conc}}

Existence of quantum Zeno and anti-Zeno effects are reported in an
asymmetric nonlinear optical coupler prepared by combing a linear
and a nonlinear waveguide of $\chi^{\left(2\right)}$ nonlinearity
(cf. Fig. \ref{fig:zeno}). Further, it is also shown that in the
stimulated case, it is easy to switch between quantum Zeno and anti-Zeno
effects just by controlling the phase of the second harmonic mode
in the asymmetric coupler discussed here. This flexibility to switch
between the quantum Zeno and anti-Zeno effects was not observed in
earlier studies on quantum Zeno and anti-Zeno effects in optical couplers\cite{thun2002zeno-raman,rehacek2000zeno-coupler,chi2-chi2-zeno}.
Further, the effects of change in linear coupling and phase mismatch
on the spatial variation of the Zeno parameter are also illustrated.
\\

The approach adopted here and in Ref. \cite{perina1995photon} is
quite general and same may be used for the similar studies on other
nonlinear optical couplers. Further, the theoretical results reported
here seem to be easily realizable in experiment as the coupler used
here is commercially available and the photon number statistics required
to study quantum Zeno and anti-Zeno effects can be obtained using
high efficiency photon number resolving detectors. 
\begin{acknowledgments}
K. T. and A. P. thank Department of Science and Technology (DST),
India for support provided through the DST project No. SR/S2/LOP-0012/2010.
They also thank B. Sen, A. Venugopalan and J. Pe$\check{{\rm r}}$ina
for some useful discussions.\end{acknowledgments}

\end{document}